\documentclass[fleqn,10pt]{wlscirep}
\usepackage{subfigure}
\usepackage{booktabs}
\usepackage{graphicx}
\usepackage{siunitx}  % for units of measure and data in tables
\usepackage{amssymb}
\usepackage{amsfonts}
\usepackage{amsmath}
\usepackage{mathrsfs}
\usepackage{algorithm} %format of the algorithm
\usepackage{algorithmic} %format of the algorithm
\usepackage{multirow} %multirow for format of table
\usepackage{amsmath}
\usepackage{xcolor}
\usepackage{cases}

\title{The Hidden Flow Structure and Metric Space of Network Embedding Algorithms Based on Random Walks}

\author[1]{Weiwei Gu}
\author[1]{Li Gong}
\author[1]{Xiaodan Lou}
\author[1,*]{Jiang Zhang}
\affil[1]{School of Systems Science, Beijing Normal University, Beijing 100875, P.R.China}
\affil[*]{zhangjiang@bnu.edu.cn}

\begin{abstract}
	Network embedding which encodes all vertices in a network as a set of numerical vectors in accordance with it's local and global structures, has drawn widespread attention. Network embedding not only learns significant features of a network, such as the clustering and linking prediction but also learns the latent vector representation of the nodes which provides theoretical support for a variety of applications, such as visualization, node classification, and recommendation. As the latest progress of the research, several algorithms based on random walks have been devised. Although their high scores for learning efficiency and accuracy have drawn much attention, there is still a lack of theoretical explanation, and the transparency of the algorithms has been doubted. Here, we propose an approach based on the open-flow network model to reveal the underlying flow structure and its hidden metric space of different random walk strategies on networks. We show that the essence of embedding based on random walks is the latent metric structure defined on the open-flow network. This not only deepens our understanding of random walk based embedding algorithms but also helps in finding new potential applications in embedding.
\end{abstract}
\begin{document}
	
	\flushbottom
	\maketitle
	
	\thispagestyle{empty}
	
	\section*{Introduction}
	Complex networks, as high-level abstractions of complex system, have been widely applied in different areas, such as biology, sociology, economics and technology\cite{RevModPhys.74, Barab2004Network, Deville2016Scaling,Marc2010Spatial,Wang2016The,Lv2009Effective}. Recent progress has revealed a hidden geometric structure in networks\cite{Brockmann2013The,Garc2016The} that not only deepens our understanding of the multiscale nature and intrinsic heterogeneity of networks but also provides a useful tool to unravel the regularity of some dynamic processes on networks \cite{Brockmann2013The,Kleinberg2000Navigation,Higham2008Fitting,Kleinberg2007Geographic,Shi2015A,Lou2016The,Serrano2012Uncovering}.
	At the same time, researchers in the machine learning community have developed several techniques to embed a whole network in a high-dimensional space\cite{Perozzi2014DeepWalk,Grover2016node2vec,Tang2015LINE,Cao2015GraRep,Arora2015RAND,Wang2016Structural} such that the vectors of each node can be used as abstract features feeding on neural networks to perform tasks. It has been demonstrated that such a form of network embedding has wide applications, such as community detection, node classification and link prediction\cite{Grover2016node2vec,Leskovec2010Empirical}. Various methods have been proposed in network embedding field such as Principal Component Analysis, Multi-Dimensional Scaling, IsoMap and their extensions\cite{Belkin2002Laplacian,Roweis2000Nonlinear,Tenenbaum2001A,Yan2005Graph,Shavitt2002Big}. Those embedding methods give good performance when the network is small. But most of them cannot be effectively applied on networks containing millions of nodes and billions of edges.
	
	Recently, there has been a surge of works proposing alternative ways to embed networks by training neural networks\cite{Perozzi2014DeepWalk,Grover2016node2vec} in various approaches inspired by natural language processing techniques\cite{Pennington2014Glove,Sridhar2015Unsupervised,Mikolov2013Efficient,Mikolov2013Distributed}. To build a connection between a language and a network, a random walk needs to be implemented on the network such that the node sequences generated by random walks are treated as sentences in which nodes resemble words. After the sequences are generated, skip-gram in word2vec\cite{Mikolov2013Distributed}, which is one of the most famous algorithms for word embedding developed in the deep learning community, can be efficiently applied on the sequences. Among these random-walk-based approaches, deepwalk\cite{Perozzi2014DeepWalk} and node2vec\cite{Grover2016node2vec} have drawn wide attention for their high training speed and high classification accuracy. Both algorithms regard random walks as a paradigmatic dynamic process on a network that can reveal both the local and global network structures. Several extended works that unravel the fundamental co-occurrence matrix between the context and words in skip-gram-based embedding algorithms and the multiple-step transition matrix. Levy et al. \cite{Levy2014Neura}proves that skip-gram models are implicitly factorizing a word-contex matrix. Tang et al.\cite{Tang2015LINE} takes 1-step and 2-step local relational co-occurrence into consideration and Cao et al.\cite{Cao2015GraRep} believes that the skip-gram is an equally weighted linear combination of k-step relational information. Those works were proposed soon after word2vec and deepwalk were presented.
	
	Although word2vec and network embedding are successfully applied in some real problems, several drawbacks still exist. First, explicit and fundamental explanations are nedded to explain why neural-based algorithms work so well since these algorithms are fundamentally black boxes. Second, how we should set the values of the hyper-parameters is still poorly understood. Third, explicit and intuitive explanations of the embedding vectors of each node and the inner structures of the embedding space are needed. We should find an explanation to provide a general framework to unify deepwalk, node2vec and other new algorithms based on random walks.
	
	In this paper, we put forward a novel framework based on an open-flow network to deepen the understanding of network embedding algorithms based on random walks. We first use the so-called open-flow network model to characterize the different random walk strategies on a single background network. Then, we note that there is a natural metric called the flow distance that is defined on these flow networks. Finally, the hidden metric space framed by the flow distances can be derived and, interestingly, this metric space is similar to the embedding space from the deepwalk and node2vec algorithms. We uncover that the embedding algorithms based on neural networks are only attempting to realize the hidden metric based on flow networks, and the correlation between flow distance and node2vec is up to 0.91. With this understanding, we propose a new method called Flow-based Geometric Embedding(FGE), which has no free parameters and performs excellently in some applications, such as clustering and node centrality ranking.
	
	\section*{Methods}
	Both deepwalk and node2vec are aim to learn the continuous feature representations of nodes by sampling truncated random walk sequences from the graph as mimic sentences to feed on the skip-gram algorithm in word2vec. The difference lies in the random walk strategy, where the deepwalk algorithm implements a common unbiased random walk on a graph such that all the edges are visited in accordance with the relative weights on the local node, while node2vec employs a biased random walk in which the probability of visiting is adjusted by two parameters $p$ and $q$. Node2vec can uncover much richer structures of a network because it resembles deepwalk when $p = 1$ and $q = 1$. Thus, we discuss only node2vec in the rest of this paper. Please reffer to algorithm \ref{alg:node2vec} for more concrete details about node2vec.
	
\subsection*{Constructing Open-flow Networks}
\begin{figure}[ht]
		\centering
		\includegraphics[width=\linewidth]{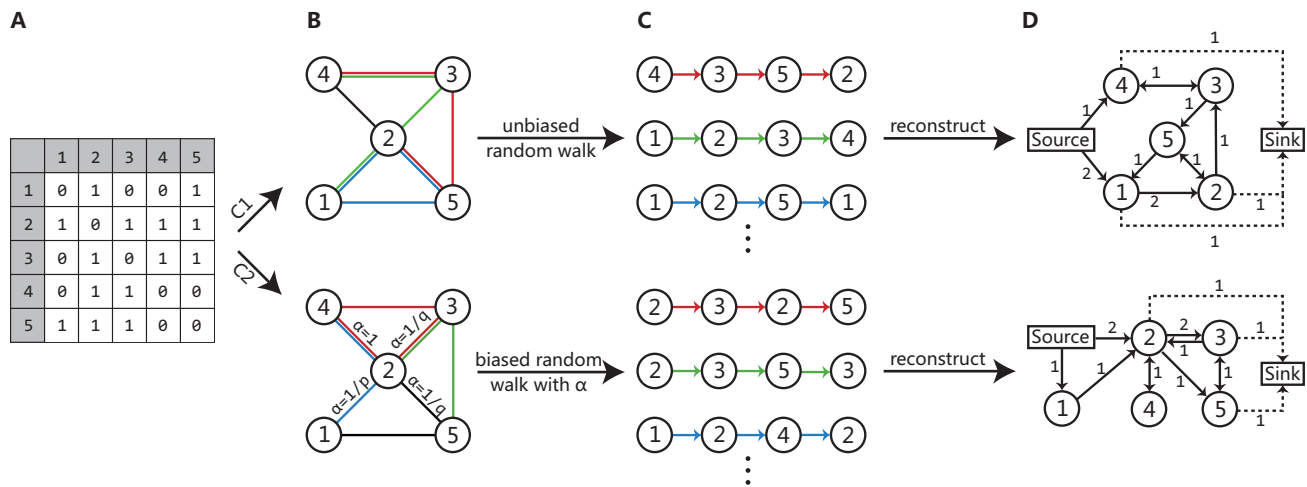}
		\caption{\textbf{Illustration of the construction of different open-flow networks from the same background network with different random walk strategies.} (A) represents the adjacency matrix of a network; (B) shows the random walks implemented by the deepwalk algorithm with $p = 1$, $q = 1$(C1), and node2vec algorithm with $p = 0.5$, $q = 1$ (C2) from (A); (C) shows several sequences of nodes generated by the corresponding random walk algorithms; (D) shows the open-flow networks constructed by the sequences. Algorithm \ref{alg:flow distance} shows how to build an open-flow network matrix based on total flow from node to node.}
		\label{fig:lou-z}
	\end{figure}

 To reveal the flow structure behind a random walk strategy (for a given $p$ and $q$), we constructed an open-flow network model\cite{Guo2015Flow} in accordance with the random walk strategy. An open-flow network is a special directed weighted network in which the nodes are identical to those of the original network, and the weighted edges represent the actual fluxes realized by a large number of random walks. There are two special nodes, the source and the sink, representing the environment, that is why the network is called an open network. When a random walker is generated at a given node, a unit of flux is injected into the flow network from the source to the given node, and this particle contributes one unit of flux to all the edges visited. When the random walk is truncated, a unit of flux is added from the last node to the sink. A large number of random walkers jumping on the network according to the specific strategy form a flow structure that can be characterized by the open-flow network model, in which the weight on the edge $i \xrightarrow{} j$ is the number of particles visited. Figure \ref{fig:lou-z} illustrates how the different open-flow networks are constructed for a single background binary network with deepwalk in the upper panel and node2vec in the lower panel.
 
\subsection*{Calculating Flow Distance}
\begin{figure}
		\centering
		\subfigure[matrix F built from random walk sequences ]{
			\label{fig:subfig:A} %% 第一幅图的标签
			\includegraphics[width=0.4\textwidth]{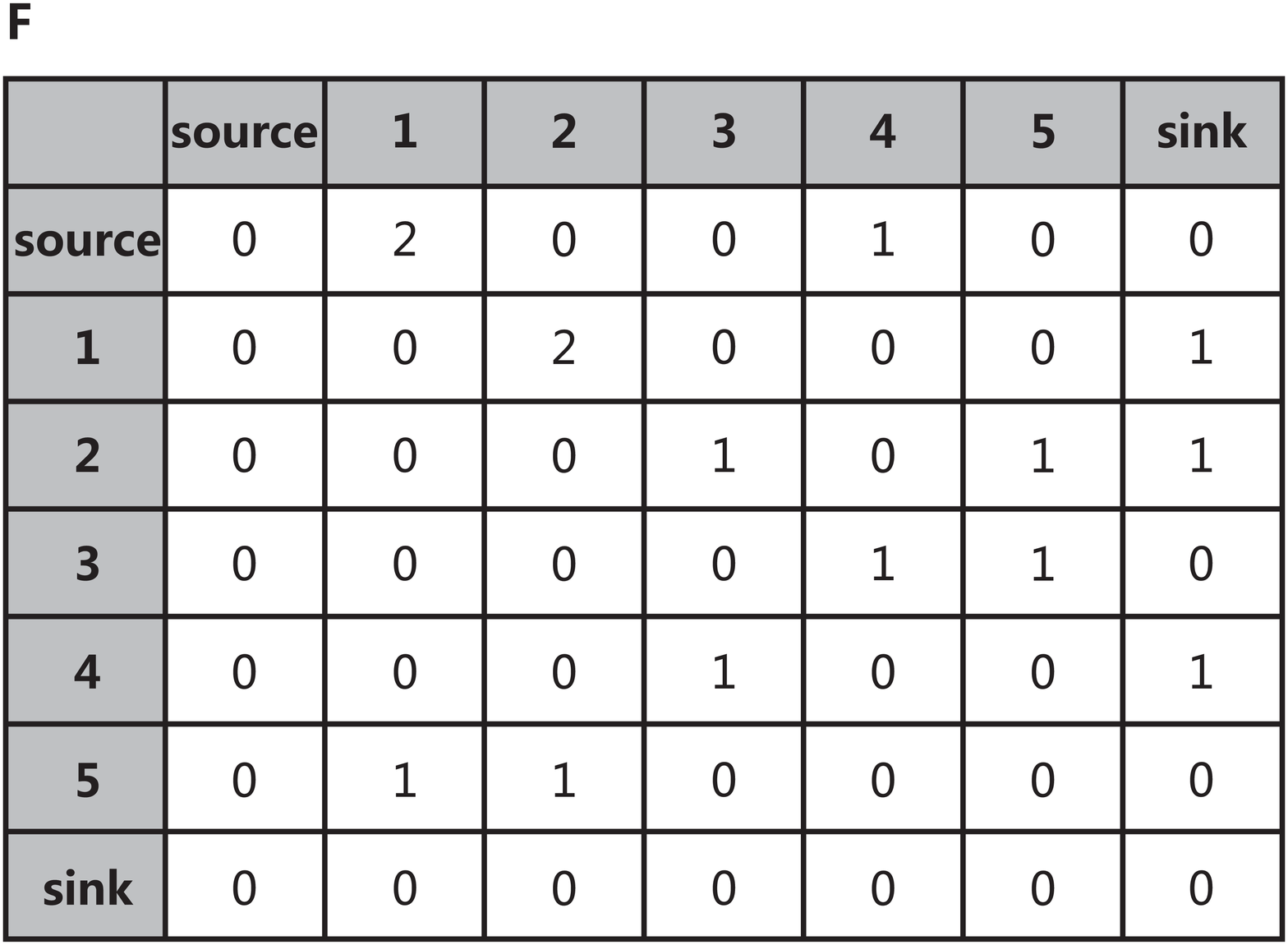}}
		\hspace{1in}
		\subfigure[matrix C calculated based on F]{
			\label{fig:subfig:B} %% 第二幅图的标签
			\includegraphics[width=0.4\textwidth]{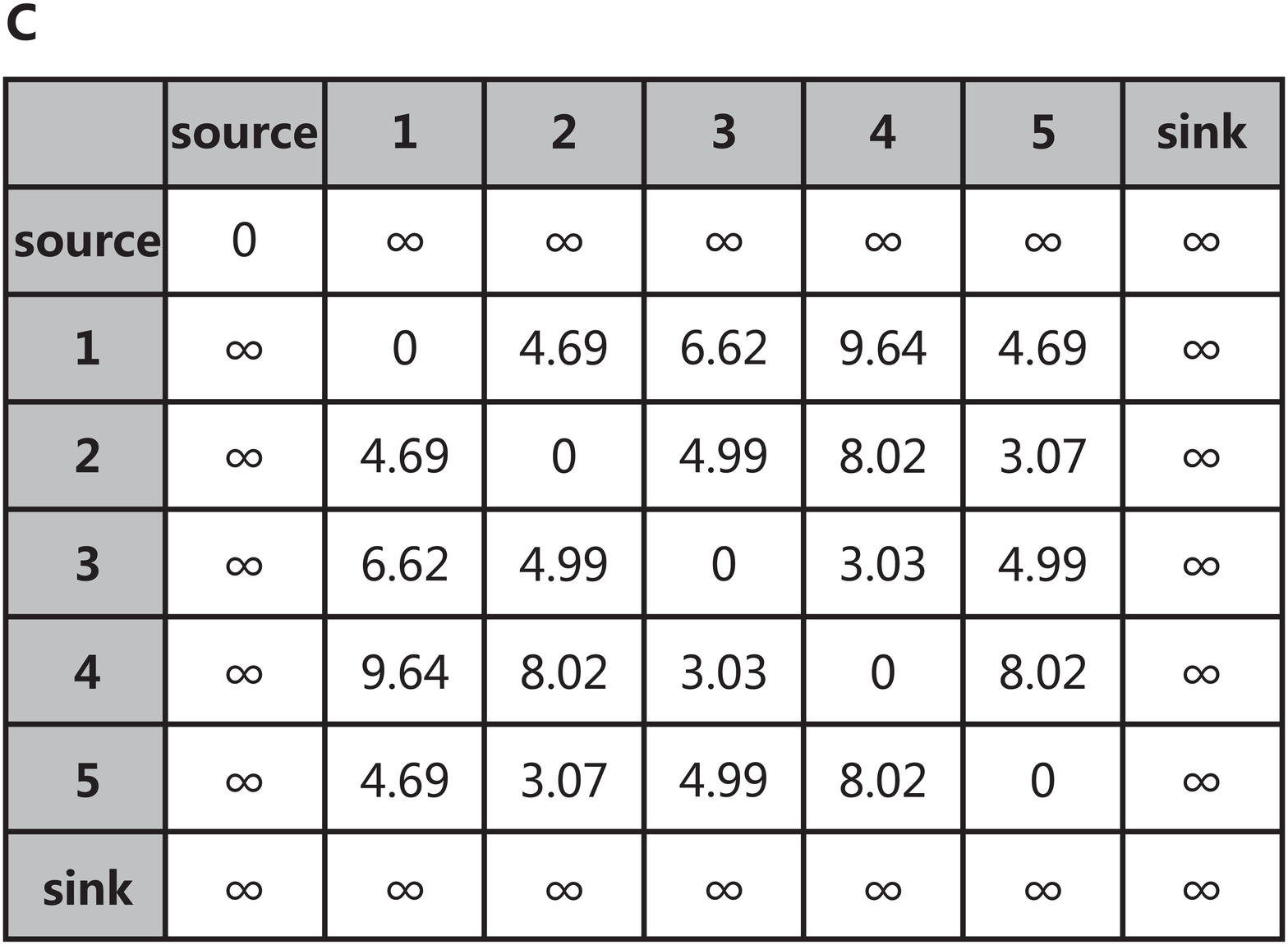}}
		\caption{\textbf{An example flow network including 7 nodes.}  (a) is the flux matrix $F$ of the sampled network under condition C1 ($p = 1$, $q = 1$) in Figure \ref{fig:lou-z}. (b) shows the flow distances among all nodes, where infinity means that there is no connected path from $i$ to $j$. Algorithm \ref{alg:flow distance} shows how to compute the flow distance based on the $F$ matrix.}
		\label{fig:entire} %% label for entire figure
	\end{figure}

For a given flow network F with $(N+2) \times (N+2)$ entries, where the value at the $i$-th row and the $j$-th column represents the flux from $i$ to $j$, node 0 represents the source, and the sink is represented as the last node, the flow distance $c_{ij}$ between any pair of nodes $i$ and $j$ is defined as the average number of steps needed for a random walker to jump from $i$ to $j$ and finally return back to $i$ along the network. It can be expressed as:
	\begin{equation}
	c_{ij} = \frac{(MU^2)_{ij}}{u_{ij}} + \frac{(MU^2)_{ji}}{u_{ji}}-\frac{(MU^2)_{ii}}{u_{ii}}-\frac{(MU^2)_{jj}}{u_{jj}}
	\end{equation}
	Where, m$_{ij}$  is the transition probability from $i$ to $j$, which is defined as:
	$m_{ij} = \frac{f_{ij}}{\sum_{j=1}^{N+1}f_{ij}} $ where $f_{ij}$  is the total flow from node $i$ to node $j$. The pseudo probability matrix $U$ is defined as\cite{Guo2015Flow}:
	\begin{equation}
	U = I + M + M^2 +...=(I-M)^{-1}
	\end{equation}
	Where $I$ is the identity matrix with $N+2$ nodes. $u_{ij}$ is the pseudo probability that a random walker jumps from $i$ to $j$ along all possible paths. Figure \ref{fig:entire} is a sample flow network constructed under condition 1 in Figure \ref{fig:lou-z}. Algorithm \ref{alg1} shows the concrete details about how to calculate flow distance based on F matrix.
	
\subsection*{Embedding Networks}
To display the hidden information in an open-flow network and visualize the node relationships, we embed the flow distance ($c_{ij}$) into a high-dimensional Euclidean space. We use the SMACOF algorithm\cite{Williams2002On,Borg2009Modern,Deville2016Scaling} to perform the embedding. This algorithm takes the distance matrix and the number of dimensions as the input and tries to place each node in $N$-dimensional space such that the between-node distance is preserved as well as possible. Through network embedding, we find the proper vector representation of $n$ nodes in the network. Please refer to algorithms \ref{alg1:embedding} for more concrete details about this embedding method.

Based on algorithms \ref{alg:flow distance} and \ref{alg1:embedding}, we proposed a new network embedding algorithm named Flow-based Geometric Embedding (FGE). We then discovered the hidden metric space of the random-walk-based network embedding algorithms, such as node2vec, word2vec GraRep and so on. The nodes’ training vectors obtained from the node2vec algorithm is highly correlated with the Euclidean distance embedding vectors derived from the flow network. The strong correlation is shown in the “Results” section.
	
	\section*{Results}
	In this section, we present our results applied on several empirical networks. First, we applied FGE algorithm on the Karate network and plotted the open-flow network models behind two different random walk strategies (with different $p$ and $q$). Next, we compared FGE algorithm with node2vec by embedding networks into two-dimensional planes. After that, we correlated the two distances, the flow distances and the Euclidean distances which is obtained from any given node pair in node2vec algorithm to show that the node2vec embedding algorithm is attempting to realize the metric of the flow distances. Then, we compared FGE and node2vec on clustering and centrality measuring tasks. Finally, we studied how the parameters of embedding algorithms based on random walks affected the flow structure and the correlations between the two distances. An overview of the networks we consider in our experiments is given in Table \ref{basic data information}.

	\subsection*{Flow Structure and Representation}
\begin{figure}[ht]
		\includegraphics[width=\linewidth]{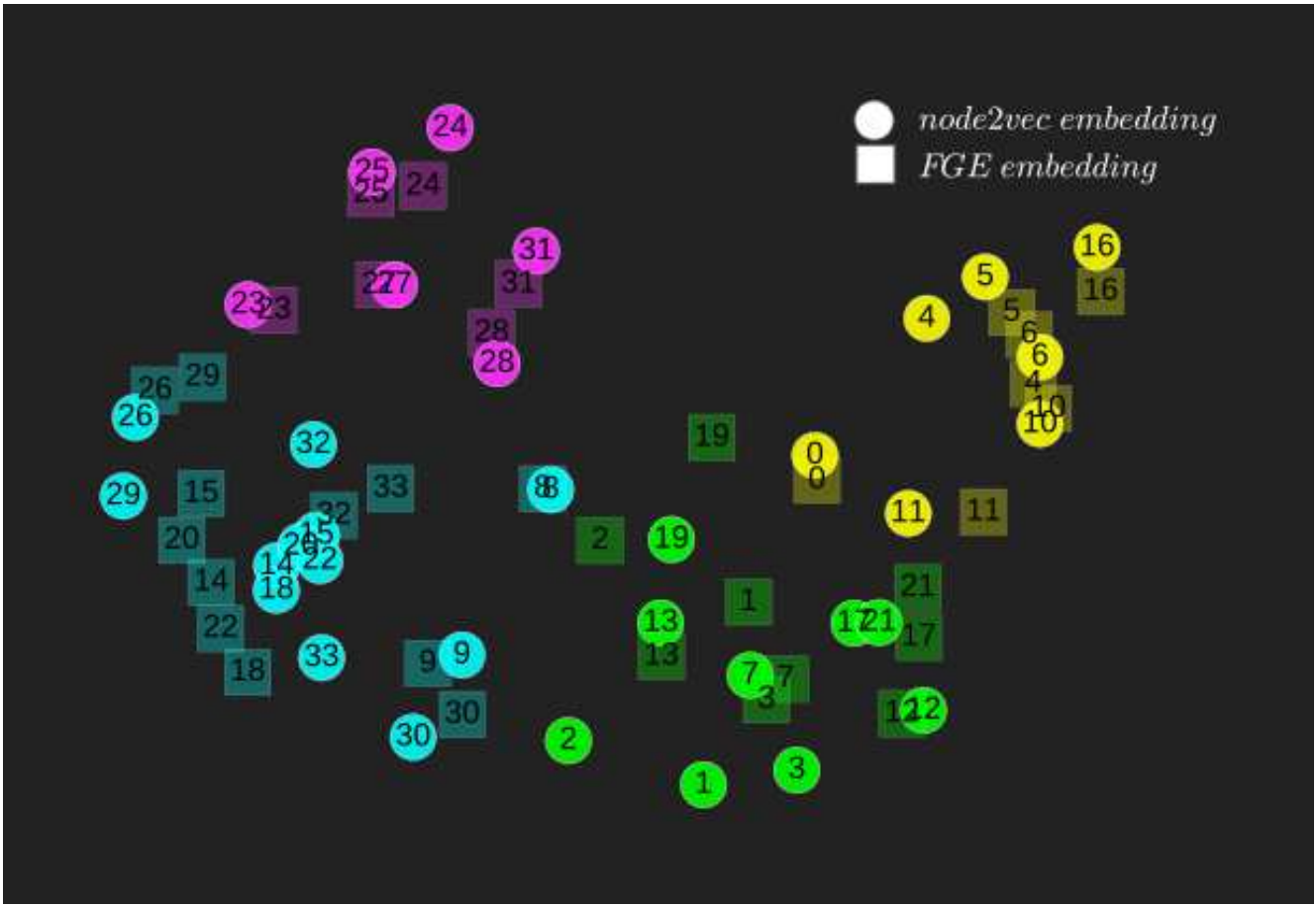}
		\caption{\textbf{The embedding of Karate Graph. }The visualization results were generated by node2vec and FGE algorithms with label colors reflecting clustering results and node shapes indicating different embedding methods.}
		\label{fig:visualization}
	\end{figure}

	This section describes experiments on Karate Graph. Figure %\ref{fig:karate} 
    shows different flow structures of node2vec with different p and q, where the thickness of the line indicates the amount of the flow between nodes. To capture the hidden metric on the flow structures, we fed random walk sequences into node2vec and FGE with  number of walks per node $r = 1024$, walk length $l = 10$ and embedding size $d = 16$. After this training process, each node could acquire two vector representations, denoted by $\theta$ in FGE and $\pi$ in node2vec. We then visualized the vector representations using t-SNE\cite{Laurens2008Visualizing}, which provided both qualitative and quantitative results for the learned vector representations. Figure \ref{fig:visualization} shows the flow structure generated by unbaised random walk strategies($p = 1$, $q = 1$).
	Figure \ref{fig:visualization} represents the visualization of $\theta$ and $\pi$ under $p = 1$ and $q = 1$. Intuitively, we observed that the nodes embedded by this two methods almost overlapped each othjavascript:void(0);er. This indicates that the flow distances of the embedding captured the essence of node2vec. Additionally, the latent relationship between nodes was well expressed. For example, we found that nodes 4,5,10 and 16 were all close to each other and belong to the same community in both algorithms. By analysing the network structure, we also discovered that nodes 14,15,20 and 22 were much closer to each other in node2vec embedding than in the FGE embedding. That is because node2vec only considers n-step connection between nodes. However, the relationship changed when we consider infinity step connection with other nodes. This change can be captured by FGE algorithm since it considers all pathways.
	
	\subsection*{Correlations between Distances}
\begin{figure}
		\centering
		\subfigure[]{
			\label{fig:subfigCorKarate} %% 第一幅图的标签
			\includegraphics[width=0.4\textwidth]{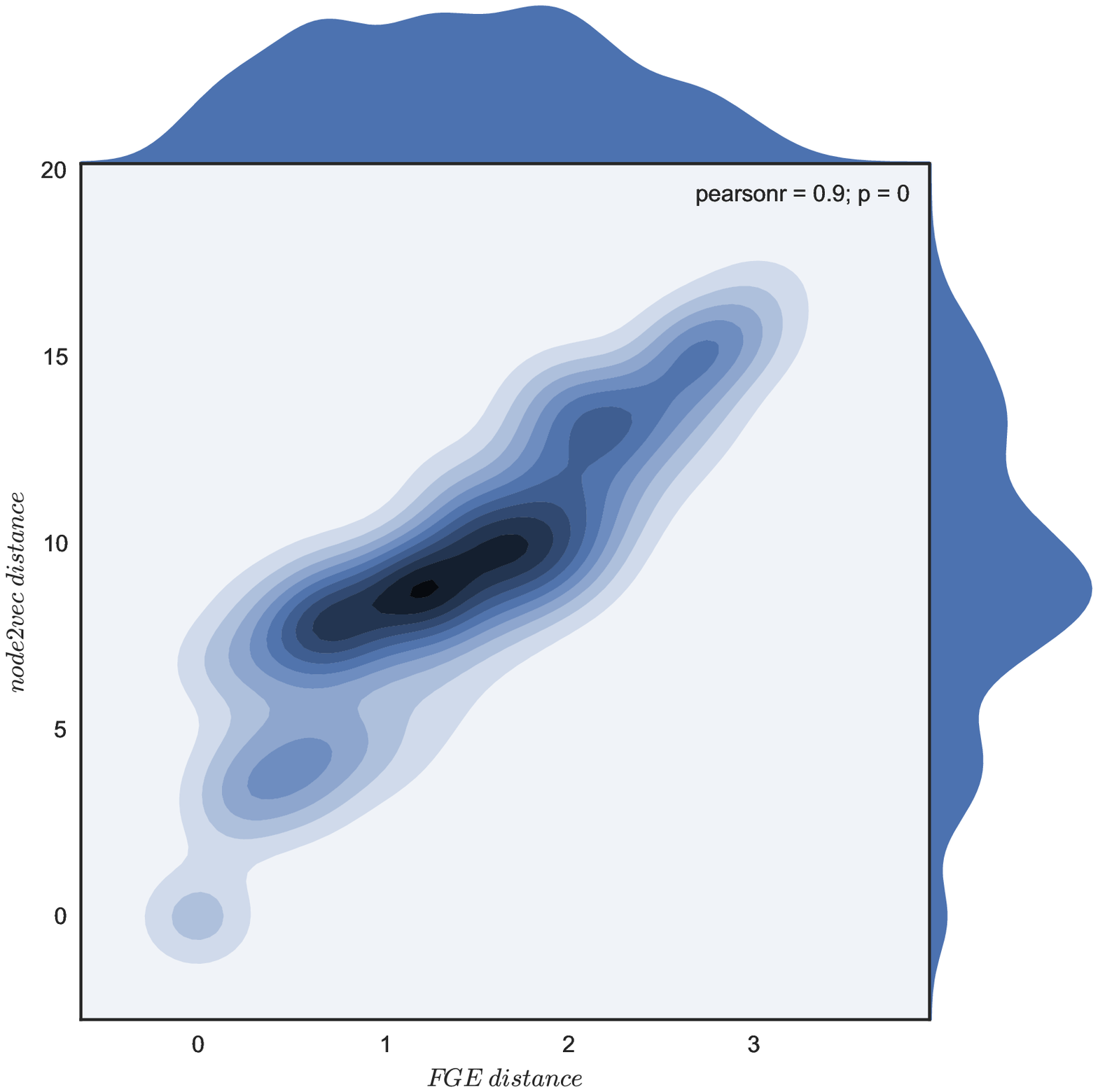}}
		\hspace{0.5cm}
		\subfigure[]{
			\label{fig:subfigCorAirline} %% 第二幅图的标签
			\includegraphics[width=0.4\textwidth]{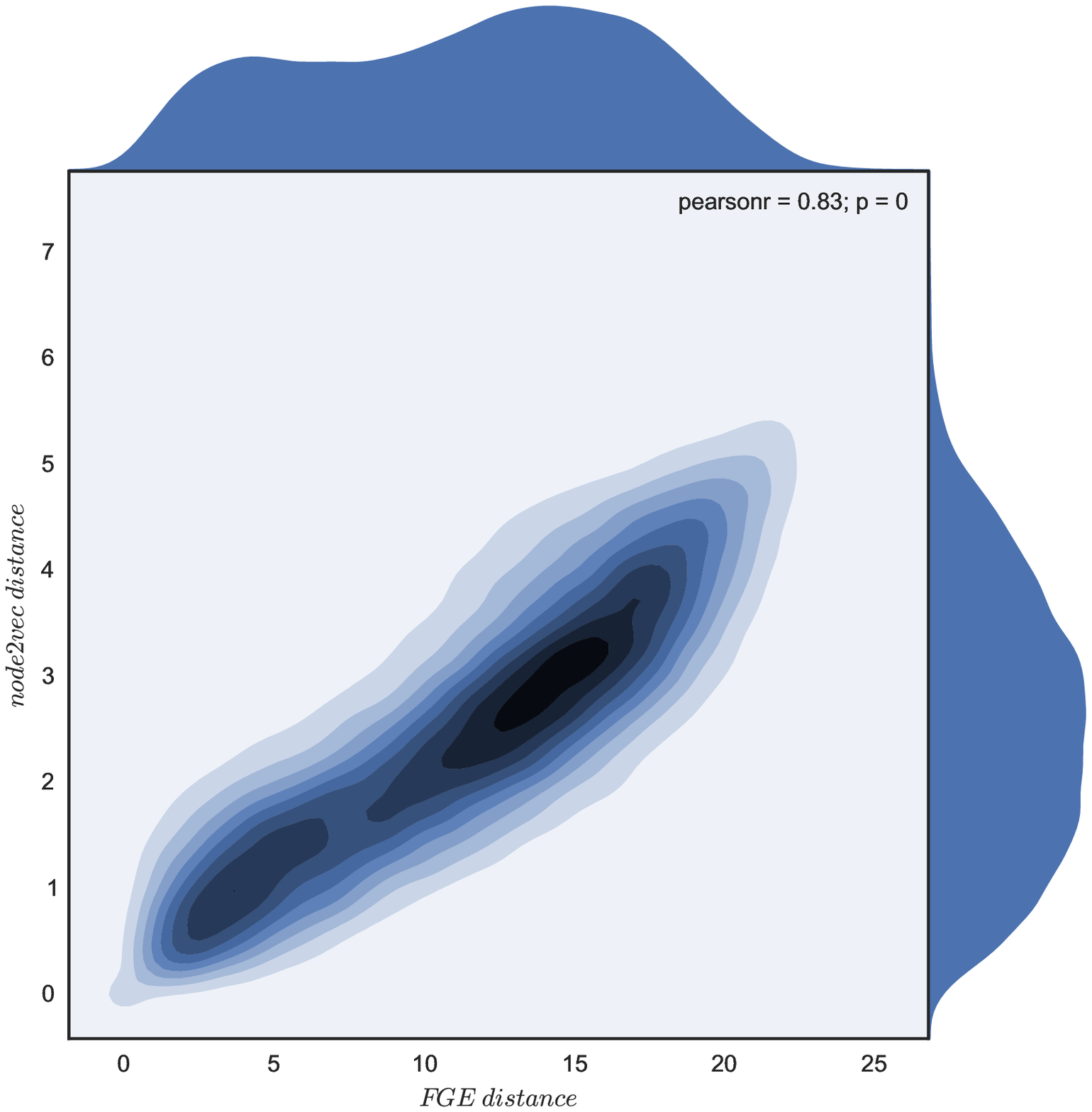}}
			\caption{\textbf{Heat maps of flow distance and node2vec distance under p=1, q=1.}(a) shows the correlation coefficient between flow distance and node2vec distance of Karate Graph. (b) indicates the correlation coefficient of Airline Network dataset.}
	
		\label{fig:correlation} %% label for entire figure
	\end{figure}

	To confirm our conclusion that the skip-gram algorithm only tries to realize the hidden metric of the flow distance defined by random walks, we plotted the flow distance of the flow network generated by random walks from FGE algorithm and the Euclidean distance in the embedding space given by the node2vec algorithm on the same node sequences for any given node pair on the same background network. The results showed strong correlations between the two distances. Figure \ref{fig:correlation} is a heat map, where the X-axis represents the flow distance. between nodes $i$ and $j$, and the Y-axis is the node2vec distance . The Pearson correlation between the two distances was 0.90 with a p-value = 0.001 in Figure \ref{fig:subfigCorKarate}  and 0.83 with a p-value = 0 in \ref{fig:subfigCorAirline}. The correlation indicates the highly linear relationship between the paired data.
	
	To show the generality of our results, we performed the same experiments over different datasets, and the accuracy of the experiments was enhanced by averaging the correlation value of each dataset. The results in table \ref{correlation information} shows that there is a strong connection between the flow distance and node2vec’s distance.
	We also found that the correlation is not sensitive to the different walking strategies. This is because different walking strategies generate different neighbor nodes, leading to different metric distances. All those walking strategies can be captured by the flow distances, and so the flow distance can reveal the latent space in random-walk-based network embedding algorithms such as node2vec,deepwalk and so on. The FGE algorithm can reveal the latent relationship between nodes in graph embedding.
	
	\subsection*{Node Clustering}
	To further show the similarity between our method and node2vec, we compared the two approaches in performing node clustering. In complex network studies, node clustering is merely community structure detection, which is of importance in various backgrounds\cite{Newman2003A,Freeman1980Centrality,Rka1999Hawoong}. We then performed the k-means clustering method on the node vectors $\mathbf{\theta}$ and $\mathbf{\pi}$ with $r=1024$, $d=2$ and $l=10$. The number of clusters can be determined using the average silhouette coefficient as the criterion. According to the silhouette value, we aggregated the graph into 4 clusters, each of them is regard as a community. Here, our method was applied to the karate club graph, as shown in Figure \ref{fig:visualization}. In this graph different vertex colors represented different communities of the input graph. The clusters obtained using the two methods overlapped in a degree of 100\%. We also performed a clustering experiment on other datasets, such as China Click Websites and Airline Network the clustering results were identical.
	
	\subsection*{Centrality Measurement}
	We showed that the understanding of the network embedding from the angle of the flow structure could not only provide us new insights but also new applications. Such as centrality measuring. The centrality measure of nodes is a key issue in network analysis \cite{Freeman1978Centrality,Bonacich1987Power,Borgatti2005Centrality} and a variety of centrality measures have been developed. Here, we showed that the average distance from the focus node to all other nodes can be treated as a new type of node centrality measure. Formally, we defined a new metric to measure the centrality of nodes based on FGE as:
	\begin{equation}
	\centering
	\bar{c_i} = \sum_{j} c_{ij}
	\end{equation}\\
	The reason for the usefulness of this definition is that the nodes close to other nodes always have tight connections and high traffic. Because the flow distances are highly correlated with the Euclidean distances in node2vec embedding, this definition also works for the node2vec algorithm. That is, we can measure each node’s centrality through its distances to all other nodes in the embedding Euclidean space. Furthermore, we can read the centrality information directly from the embedding graph because the nodes with high centrality (small average distances) are always concentrated on the central area of the embedding graph.
	
	We tested the node centrality on the dataset of China Click Websites, which contained approximately 5 years of browsing data from more than 30000 online volunteers. We calculated each website’s centrality based on its flow distance matrix and node2vec distance. We found that the popular websites have always had a small distance because they usually have had more travelling paths to other websites. Therefore, the smaller the average flow distance, the more central the website position. We ranked the websites in accordance  with their centrality and then compared those two methods with other methods such as PageRank and total traffic (the number of clicks for each website). The ranking results for the top 10 websites are listed in Table \ref{table: ranking}. We found that the ranking orders of the flow distance and node2vec were nearly the same.
	We also discovered that high-traffic websites, such as Tmall (a popular shopping website) and 163.com, have lower ranks, but baidu.com and qq.com have high ranking orders even though their total traffic was not heavy. That is because baidu.com and qq.com are bridges between the real and virtual worlds.
	
	\subsection*{Parameter Sensitivity} 
\begin{figure}
		\centering
		\subfigure[]{
			\label{fig:Parameter Sensitive} %% 第一幅图的标签
			\includegraphics[width=0.4\textwidth]{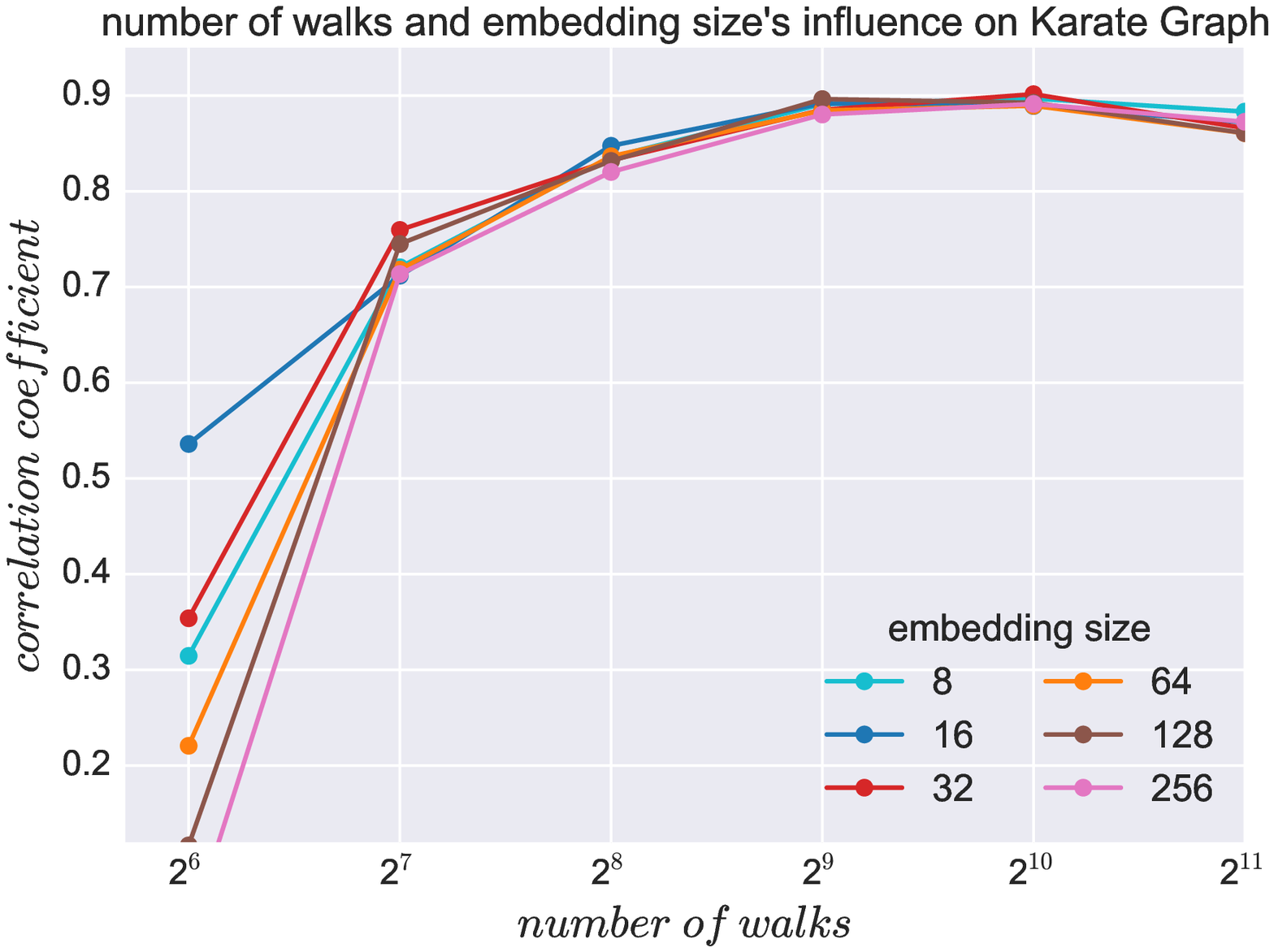}}
		\hspace{0.1cm}
		\subfigure[]{
			\label{fig:Winsize} %% 第二幅图的标签
			\includegraphics[width=0.4\textwidth]{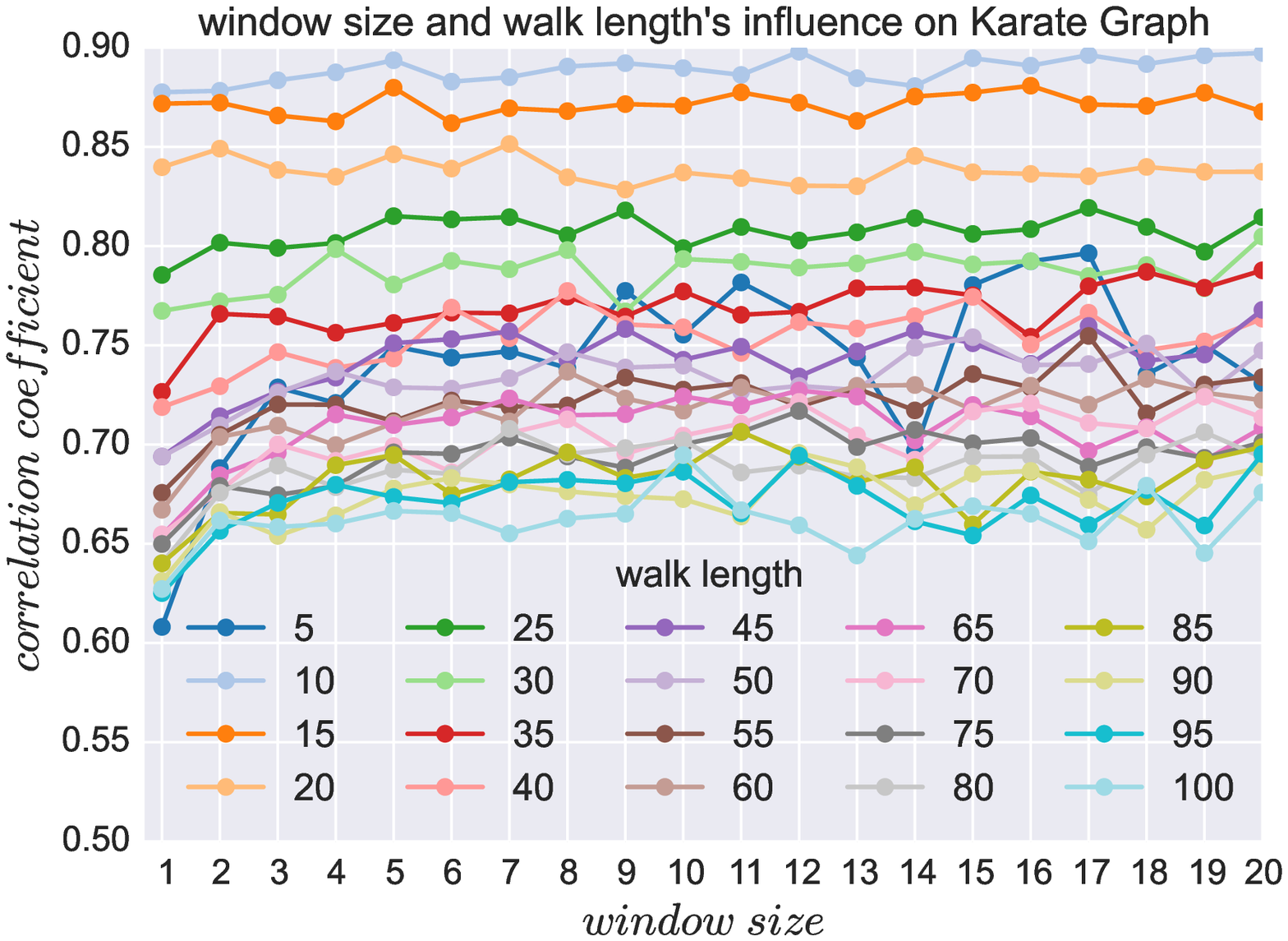}}
		\caption{\textbf{Parameter Sensitivity Study.} (a) The correlation coefficient over embedding size $d$ and number of walks per node $r$. (b) The correlation coefficient over walk length $l$  and number of walks per node $r$.}
		\label{fig:Sensitive} %% label for entire figure
	\end{figure}

	Random-walk-based embedding algorithms involve a number of sensitive parameters. To evaluate how the parameters affect the correlation between the two distances, we conducted several experiments on the dataset of the Karate club network. We examined how the embedding size $d$, the number of walks started per node $r$, the window size $w$, and the walk length $l$ influenced the correlation between the two distances. As shown in Fig \ref{fig:Parameter Sensitive}, the correlation grew with the  number of walks increased, and the correlation tended to saturate when the number reached 512. This indicated that the node2vec embedding algorithm merely tried to realize the hidden metric of the flow structure of the random walk, and the performance increased as more samples were drawn. The neural network of the skip-gram algorithm behind node2vec is over-fitted when the number of walks is small because a higher embedding size $d$ leads to more parameters in the neural network that needed to be trained and the correlation decreased with the embedding size $d$ (Figure \ref{fig:Parameter Sensitive}). However, there was a slight trend of the decreasing correlation coefficient with the number of walks when this number is larger than 512. We speculated that the decrease in the correlation is due to errors in the substitution of the large sample of random walks using the open-flow network.  The FGE algorithm assumes that the random walks can be represented as a Markovian process on the network, which means that each step jump is exclusively determined by the previous-step position. However, the random walk of node2vec does not satisfy this condition. Even though the difference exists as seen in Figure \ref{fig:Winsize} we believe that the hidden metric of flows is more essential to reflect the structural properties of the network.
	We also evaluated how changes to the window size $w$ and walk length $l$ affected the correlation.  We have fixed the embedding size and the number of walks to sensible values $d = 128$, $r = 512$ and varied the window size $w$ and walk length $l$ for each node. The performance differences were not that large as $w$ changed. When the walk length $l$ reached $10$, the correlation declined rapidly with further increases in the walk length.
	
	\section*{Conclusions and Discussions}
	In this paper, we reveal the hidden flow structure and metric space of random-walk-based network embedding algorithms by introducing FGE algorithm. This algorithm takes the flow from node to node as an input. After calculating the flow distance, node2vec learns nodes representation that encodes both structural and local regularities. The high Pearson correlation value between the node2vec representations and FGE vectors indicates that there is a hidden metric of random-walk-based network-embedding algorithms. The FGE algorithm not only helps in finding the hidden metric space but also works as a novel approach to learn the latent relations between vertices. Experiments on a variety of different networks illustrate the effectiveness of this method in revealing the hidden metric space of random-walk-based network-embedding algorithms.
	This finding is of great importance because it not only provides a novel perspective to understand the essence of network embedding based on random walks but also reveals the skip-gram (the main algorithm in node2vec) is trying to find the proper node representation to match this metric between nodes. With this finding, we first applied node2vec to a centrality measuring task we use the Euclidean distance instead of cosine distance between nodes to measure the importance of nodes. We then validate the Euclidean distance of the nodes’ vectors in FGE and node2vec in clustering task. The outcome shows that the two algorithms give similar clustering and centrality measuring results. The FGE algorithm has no free parameters, so it can work as a criterion for parameter setting for node2vec.  PPMI \cite{Levy2014Neura} shows that the skip-gram in word2vec is implicitly factorizes a word-context matrix. In future, we would like to explore the hidden relationship between the flow distance and point wise mutual information. Both node2vec and FGE regard random walk as a paradigmatic dynamic process to reveal network structures. This sampling strategy consumes a large amount of computer resources to reach a stationary state for each node. Further extensions of FGE could involve calculating the nodes’ flow distances without sampling.

	\bibliography{sample}
	
	\section*{Acknowledgements}
	We acknowledge the financial support for this work from the National Science Foundation of China with grant number 61673070, "the Fundamental Research Funds for the Central Universities", grant number 310421103, and Beijing Normal University Interdisciplinary Project.
	
	\section*{Author contributions statement}
	
	JZ and WG conceived the experiments and write the manuscript. WG collects and analyses the empirical data. WG, LG and XL plot the graphs. All authors reviewed the manuscript.
	
	\section*{Additional information}
	\noindent \textbf{Competing financial interests:}   The authors declare no competing financial interests.

	\begin{table}[htbp]
		\centering
		
		\begin{tabular}{l c c c }
			\toprule
			NAME & NODES & EDGES & DIRECTED \\
			\midrule
			Les Misérables network & 77	  & 254	    & directed \\
			Airline Network		 & 3,425  & 67,333  & directed \\
			Karate Graph			 & 34     & 156     & undirected \\
			Blog Catalog           & 10,312 & 333,983 & undirected \\
			Wikipedia	             & 9488	  & 832,408 & directed weighted \\
			China Click Websites   & 20,746 & 135,770	& directed weighted\\
			\bottomrule
		\end{tabular}
		\caption{\label{basic data information} \textbf{An overview of the basic information of the datasets.}  Les Misérables is a co-occurrence network with 77 characters and 254 relationships in the novel Les Misérables.
			Airline Network contains 59036 routes between 3209 airports on 531 airlines spanning the globe, the routes are directional. Karate Graph is a social network of friendships between 34 members of a karate club at a US university.
			Blog Catalog is a network of 333,983 social interactions of 10,312 bloggers listed on the Blog Catalogue website.
			Wikipedia datasets comes from the latest web page texts from the Chinese Wikipedia with 9488 nodes and 832,408 edges.
			China Click Websites contains 120 million records of all the clicking behaviours of 1000 users with 20,746 websites and 135,770 click streams within one month.
		}
	\end{table}
	\vbox{}

 	\begin{figure}[ht]
 		\centering
		\includegraphics[width=\linewidth]{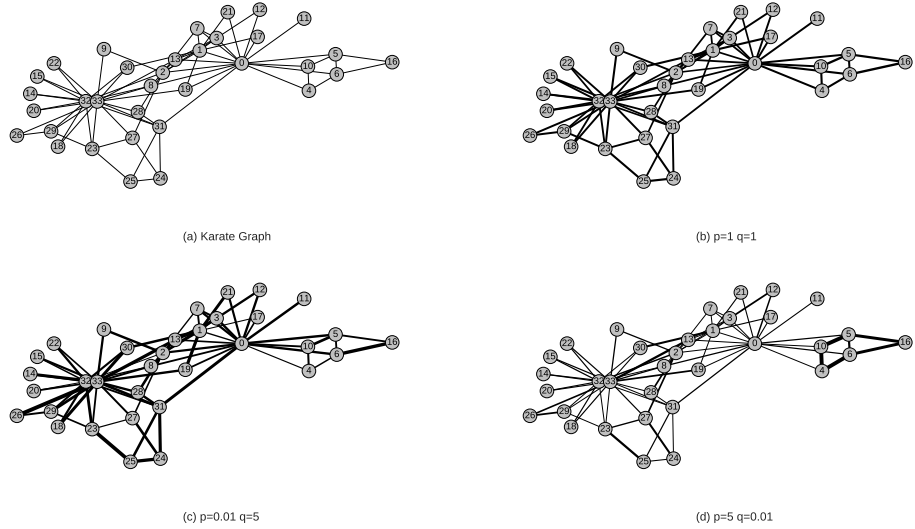}
		\caption{ \textbf{Visualization of flow network structure.} (a) is an undirected, unweighted network of Karate Graph. (b) is the result of unbaised random walks on this graph. (c) indicates the random walks mainly explored within a community to uncover the local structures, while in (d) the vast flows between different communities showed the random walks were trying to find the global structures in this network. }
 		\label{fig:karate}
 	\end{figure}
	
	\begin{table}[htbp]
		\centering
		\begin{tabular}{l c c c c c c}
			\toprule
			Parameter & Karate Graph & Blog Catalog & Les Misérables Network & Airline Network & China Click Websites & Wikipedia \\
			\midrule
			p=1.0, q=1.0 & 0.91	& 0.78	& 0.81 & 0.82 & 0.86 & 0.58\\
			p=0.5, q=2.0 & 0.90	& 0.86	& 0.74 & 0.85 & 0.60 & 0.62\\
			p=2.0, q=0.5 & 0.87	& 0.79	& 0.86 & 0.77 & 0.75 & 0.65\\
			\bottomrule
		\end{tabular}
		\caption{\label{correlation information} \textbf{The correlation coefficients of different datasets.} This table shows the coefficients between flow distance and node2vec’s Euclidean distance in embedding space, in which node2vec is chosen with different “inward” and “outward” parameters ($p$ and $q$).
		}
	\end{table}

	\begin{table}[htbp]
		\centering
		\begin{tabular}{l l c c c c }
			\toprule
			rank & web name & flow distance &  node2vec distacne & PageRank & total flow \\
			\midrule
			1  & baidu  & (1)26.332	& (1)26.437 & (1)0.0221	& (1)105560 \\
			2  & qq	    & (2)30.087	& (2)30.208	& (2)0.0189	& (2)57209  \\
			3  & sogou  & (3)33.035 & (4)33.168 & (3)0.0131	& (4)25979  \\
			4  & taobao & (4)33.272 & (3)33.106 & (4)0.0120	& (3)35311  \\
			5  & hao123 & (5)33.626 & (5)34.190 & (6)0.0122	& (5)23295  \\
			6  & sina	& (6)33.818 & (7)33.954 & (5)0.0098	& (7)21711  \\
			7  & weibo	& (7)34.054 & (6)33.953 & (9)0.0070	& (6)21815  \\
			8  & 163	& (8)34.949 & (12)36.41 & (7)0.0062 & (12)13890 \\
			9  & sohu   & (9)35.706	& (8)33.239 & (8)0.0071 & (8)15512  \\
			10 & 360	& (10)35.01 & (9)35.155 & (10)0.006	& (9)14711  \\
			
			\bottomrule
		\end{tabular}
		\caption{\label{table: ranking} \textbf{Centrality ranking of top 10 websites.}  Ranking top 10 websites according to flow distance, node2vec distance and comparisons with other ranking methods.
		}
	\end{table}
	
	\renewcommand{\algorithmicrequire}{\textbf{Input:}}
	\renewcommand{\algorithmicensure}{\textbf{Output:}}
	\begin{algorithm} %算法开始
		\centering
		\caption{Flow Distance($F$, $n$)}\label{alg:flow distance} %算法的题目
		\label{alg1} %算法的标签
		\begin{algorithmic}[1] %此处的[1]控制一下算法中的每句前面都有标号
			\REQUIRE  \ total total flow from node to node matrix \quad $F$\\
			\setlength{\parindent}{2em}	number of nodes \quad $N$
			\ENSURE flow distance matrix \quad $C$ \\
			
			\STATE \textbf{Build the transition matrix based on:} $F$
			\FOR{$i=0$ to $N$}
			\FOR{$j=0$ to $N$}
			\STATE $m_{ij} = \frac{f_{ij}}{\sum_{j=0}^{N+1}f_{ij}}$
			\ENDFOR
			\ENDFOR
			\STATE \textbf{Calculate fundamental matrix U based on M and it's identity matrix:} $I$
			\STATE $U = (I-M)^{-1}$
			\FOR{$i=0$ to $N$}
			\FOR{$j=i$ to $N$}
			\STATE   $l_{ij} = \frac{(MU^2)_{ij}}{u_{ij}} -\frac{(MU^2)_{jj}}{u_{jj}}$
			\noindent{	\STATE \textbf{Symmetrize flow distance matrix:}} $C$	
			\STATE	 $c_{ij}=c_{ji} = l_{ij}+l_{ji}$
			\ENDFOR
			\ENDFOR
			
		\end{algorithmic}
	\end{algorithm}
		
	\renewcommand{\algorithmicrequire}{\textbf{Input:}}
	\renewcommand{\algorithmicensure}{\textbf{Output:}}
	\begin{algorithm} %算法开始
		\centering
		\caption{Network Embedding($C$, $d$, $iter$, $eps$, $n$)}\label{alg:network embedding} %算法的题目
		\label{alg1:embedding} %算法的标签
		\begin{algorithmic}[1] %此处的[1]控制一下算法中的每句前面都有标号
			\REQUIRE  \ flow distance matrix \quad $C$\\
			\setlength{\parindent}{2em}	embedding size \quad $d$\\
			\setlength{\parindent}{2em}	maximum number of iterations \quad $iter$\\
			\setlength{\parindent}{2em}	tolerance error to declare converge \quad $eps$\\
			\setlength{\parindent}{2em}	number of nodes for embedding \quad $n$\\
			
			\ENSURE  matrix of nodes representation  \quad $X$ $\in$ $\mathbb{R}^{n \times d}$
			\STATE \textbf{Initialization:}
			\STATE Set $Z = X^{[0]}$, $k = 0$, where $X^{[0]}$ is random started configuration, $k$ is the counter for iteration.
			\STATE compute $\sigma$(X$^{[0]}$) $= {\displaystyle{\sum_{i<j}(c_{ij}-d_{ij}(X^{[0]})})^2} $, where d$_{ij}$is the Euclidean distance between node $i$ and $j$ in $X$.
			%	$d_{ij}(X$^{[0]}$)}
			\WHILE{$k < iter \ or $ \ $\sigma$(X$^{[k-1]}$) -($\sigma$(X$^{[k]}$) $>$ $eps$  }
			\STATE $k = k+1$
			\STATE \textbf{Compute the Guttman transform:} X$^{[k]}$
			\STATE $X^{[k]} = n^{-1}B(Z)Z$,\ where:
			\begin{numcases}
			{b_{ij}=}\nonumber
			-\frac{c_{ij}}{d_{ij}(Z)}, & if $d_{ij}(Z) \neq 0$ , $i \neq j$\\ \nonumber
			0, & if $d_{ij}(Z) = 0$ , $i \neq j$\\ \nonumber
			-\displaystyle{\sum_{j=1,j \neq i}b_{ij}}, & if $i =j$\\ \nonumber
			\end{numcases}
			\STATE \textbf{Update stress function:}
			\STATE 	 $\sigma$(X$^{[k]}$) $ =  {\displaystyle{\sum_{i<j }(c_{ij})-d_{ij}(X^{[k]})})^2} $
			\STATE   $Z = X^{[k]}$
			\ENDWHILE	
		\end{algorithmic}
	\end{algorithm}

	\renewcommand{\algorithmicrequire}{\textbf{Input:}}
	\renewcommand{\algorithmicensure}{\textbf{Output:}} 		
	\begin{algorithm} %算法开始
		\centering
		\caption{Node2vec Embedding($G$, $w$, $d$, $r$, $t$, $p$, $q$)}\label{alg:node2vec} %算法的题目
		\label{algNode2vec} %算法的标签
		\begin{algorithmic}[1] %此处的[1]控制一下算法中的每句前面都有标号
			\REQUIRE  \ graph $G$($V$, $E$, $W$) \quad \\
			\setlength{\parindent}{2em}	window size \quad $w$\\
			\setlength{\parindent}{2em}	embedding size \quad $d$\\
			\setlength{\parindent}{2em}	walks per vertex \quad $r$\\
			\setlength{\parindent}{2em}	walk length \quad $l$\\
			\setlength{\parindent}{2em}	in and out parameter \quad $p$, $q$\\
			\ENSURE  matrix of nodes representation  \quad $\Theta$ $\in$ $\mathbb{R}^{n\times d}$
			\STATE \textbf{Modify Graph Weight:}
			\STATE $\pi$ = PreprocessModifiedWeights($G$, $p$, $q$)
			\STATE $G^{'}$ = ($V$, $E$, $\pi$)
			\STATE \textbf{Generate Walking Sequences:}
			\FOR{$i=0$ to $r$}
			\STATE $\Omega$ = Shuffle($V$)
			\FOR{each $v_{i}$ $\in$ $\Omega$}
			\STATE   $\Omega_{v_{i}}$ = RandomWalk($G$, $v_{i}$, $t$)
			\ENDFOR  \\
			
			\ENDFOR
			\STATE \textbf{Learn Features by SkipGram:}
			\FOR{each v$_{j}$  $\in$ $\Omega_{v_{i}}$ }
			
			\FOR{each u$_{k}$ $\in$ $\Omega_{v_{i}}[j-w:j+w]$}
			\STATE  $J(\Theta) = -logPr(u_{k}|\Theta(v_{j}))$
			
			\STATE	$ \Theta = \Theta - $$\alpha$$* \frac{\partial J}{\partial \Theta} $
			\ENDFOR
			
			\ENDFOR
			
		\end{algorithmic}
	\end{algorithm}

\end{document}